\documentclass[a4paper,11pt]{article}
\usepackage{jinstpub} % for details on the use of the package, please see the JINST-author-manual
\usepackage{lineno}
\usepackage{graphicx}
\usepackage{subcaption}

\title{Cryogenic setup for the characterization of wavelength-shifting materials for noble element radiation detectors}

\author[a]{S. Choudhary}
\author[a,1]{A.\,F.\,V. Cortez\note{Corresponding author.}}
\author[a]{M. Ku\'zniak}
\author[a]{G. Nieradka}
\author[a]{T.~Sworobowicz}
\affiliation[a]{AstroCeNT, Nicolaus Copernicus Astronomical Center of the Polish Academy of Sciences,\\ ul. Rektorska 4, 00-614 Warsaw, Poland}

\author[b]{Ł.~Świderski}
\author[b]{and T.~Szczęśniak}
\affiliation[b]{National Center for Nuclear Research,\\
05-400 Otwock - Świerk, Poland}
\emailAdd{acortez@camk.edu.pl}

\abstract{In the present work, we describe a cryogenic setup for studies of wavelength-shifting materials for optimised light collection in noble element radiation detectors, and discuss the commissioning results. This SiPM-based setup uses alpha induced scintillation in gaseous argon as the vacuum ultraviolet light source with the goal of characterising materials, such as polyethylene naphthalate (PEN) and tetraphenyl butadiene (TPB), in terms of their wavelength-shifting efficiency. Further extensions of the system are currently being studied. The foreseen upgrades are expected to allow the study of GEM-like structures potentially interesting for rare-event searches. The design of the setup will be addressed along with the first results.}

\keywords{dark matter detectors; gaseous detectors; micropattern gaseous detectors; wavelength shifters; noble gas detectors; noble liquid detectors; time projection chambers; FAT-GEM}

\begin{document}
\maketitle
\flushbottom

\section{Introduction}\label{sec:intro}
Compelling astrophysical and cosmological evidence for the existence of dark matter (DM) has led to numerous direct detection experiments, including DarkSide, XENON, LZ, etc., searching for particle DM candidates~\cite{aprile2006, dmreview, darkside, xenon, lz}. These experiments rely on noble liquid detectors, sensitive to vacuum ultraviolet (VUV) scintillation or scintillation and ionization induced by elastic scattering of WIMPs on nuclei, currently setting the most stringent upper limits for direct DM searches in the 1-10$^3$~GeV/$c^2$ WIMP mass range.

In experiments relying on dual-phase Time Projection Chambers (TPCs), ionization is detected through proportional scintillation of electrons drifted into the gas pocket on top of the detector, i.e. the secondary scintillation, S2. Both S2 and the primary scintillation signal, S1, are detected directly or after conversion to visible light. Normally, a combined analysis of both signals is used to perform event reconstruction and particle identification.

One of the main challenges in argon-based detectors is the relatively low efficiency of available VUV-optimized photosensors, currently not exceeding 15\% ~\cite{vuvsipms}. This limitation makes light collection and detection of S1 and S2 light in liquid argon (LAr) challenging. Therefore, efficient wavelength shifter (WLS) materials~\cite{wlsreview} are needed to enable light collection with standard blue-sensitive photosensors such as conventional photomultiplier tubes (PMT), silicon photomultipliers (SiPM) or avalanche photodiodes (APD).

The last decade has witnessed substantial progress in understanding the behavior, performance, and stability of commonly used WLS compounds, as well as in the development of new methods of applying WLS to the development of new structures capable of enhancing scintillation light detection.
As future experiments require much larger target masses (multi-ton scale) to improve current sensitivity limits, new WLS technologies scalable to such sizes are mandatory to improve or even maintain other critical design parameters, e.g. energy threshold or resolution.
To this end, there have been  exciting developments both in identifying new WLS compounds and solutions, as well as innovative applications of known WLS ~\cite{wlsreview}. An example, scalable to very large surface areas, are the polyethylene naphthalate (PEN) wavelength shifter or the FAT-GEM concept (field-assisted transparent gaseous electroluminescence multiplier)~\cite{fatgem}, combined with reflecting and WLS coatings~\cite{wlsfatgem} to maximize S2 light collection.

Since the specific requirements of each application can determine a different optimal WLS, there is no single universal ideal for all applications. This work aims to provide a setup for characterizing wavelength-shifting materials and structures to optimize light collection in noble element radiation detectors.

\section{ArGSet – a cryogenic Argon Gas Setup for wavelength-shifter testing at Astrocent}\label{sec:xxx2}
In order to characterize the wavelength-shifting capabilities of new materials a dedicated cryogenic setup, ArGSet was recently commissioned and tested at Astrocent. It was designed to provide information on:
(i) Efficiency and time response of WLS materials;	(ii)  Temperature dependence of WLS materials;
(iii) Compatibility of materials with cryogenic operation;
(iv) SiPM performance in cryogenic conditions.

\begin{figure}[h!] 
	\begin{center}  
	\includegraphics[height=3in,width=3in]{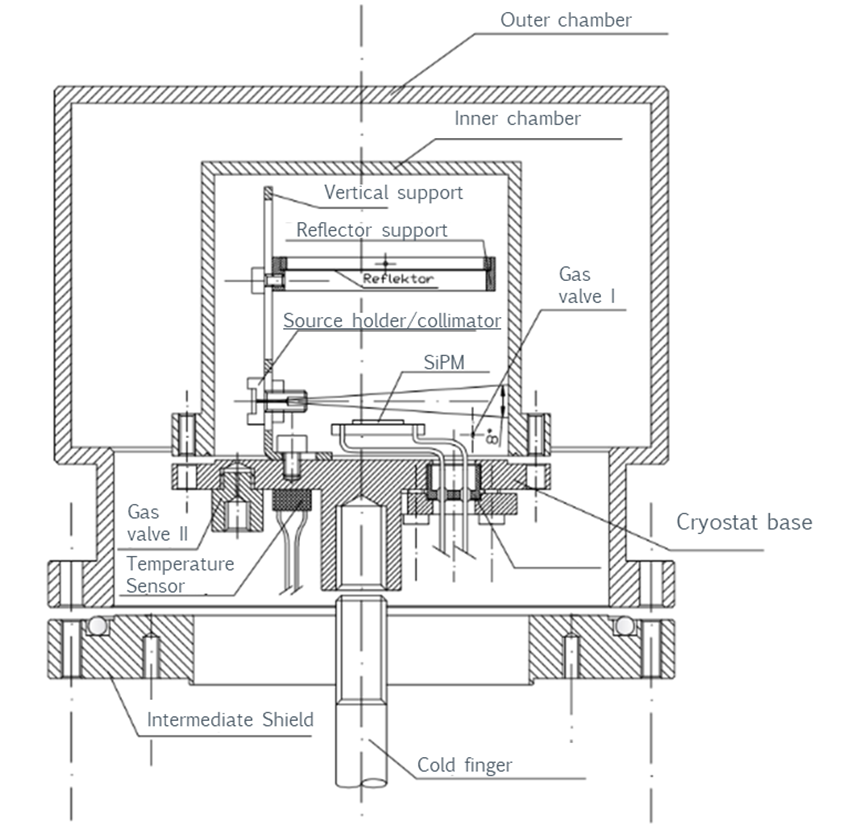}\includegraphics[height=2.5in,width=3in]{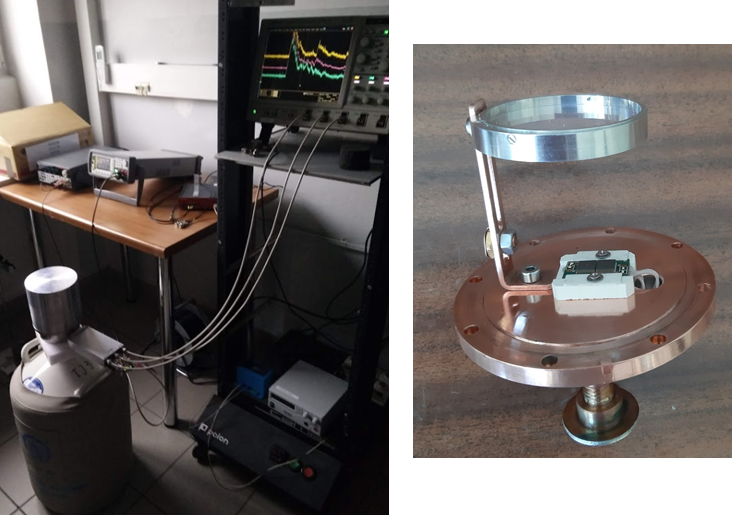}
		\caption{(Left) Schematic of the cryogenic setup for wavelength-shifter testing - ArGSet. (Middle) ArGSet experimental setup during data taking. (Right) Detailed photo of the internal structure of the cryogenic setup with the boron nitride sample holder visible at the top and the 2 SiPMs and respective support at the bottom.}\label{fig:xx4}  
	\end{center}  
\end{figure}

The schematic of the cryostat recently commissioned at Astrocent can be seen in Fig.~\ref{fig:xx4}\,(left).

The cryostat setup consists of two concentric aluminum chambers: an outer vacuum chamber that provides the thermal insulation and an indium-sealed inner chamber containing a sample holder/reflector, an $^{241}$Am $\alpha$ source holder with a collimator and two SiPMs (Hamamatsu S14160-6050HS) that can be biased and read independently. 
In order to enable $\cal O$(ns) timing-resolution for the characterization of the WLS materials, primary scintillation (S1) from Ar is used as source. To produce a measurable signal, the system relies on $^{241}$Am $\alpha$ particles as mentioned before. As a result of the interaction of the $\alpha$ particles with the Ar atoms in the inner chamber, a large number of de-excitation VUV photons will be produced. The VUV photons emitted will impinge on the sample material (WLS material to be tested), with the resulting WLS light being detected by a set of 2 independent SiPMs, not sensitive to the VUV emission spectra of Ar (1$^{st}$ and 2$^{nd}$ continuum). The distances are optimized to place the Bragg energy deposition maximum between the SiPMs and the sample.

For efficient thermal contact, SiPMs are coupled to the copper bottom plate with a boron nitride holder, while the sample is held in a copper support structure, see Fig.~\ref{fig:xx4}\,(right). The bottom plate has a thermometer attached and itself acts as a cold finger. It is directly coupled to a vacuum-jacketed copper rod, with non-insulated tip immersed in a 15L liquid nitrogen dewar, inspired by previous work~\cite{ncbj}. 
SiPM signals are routed with vacuum feedthroughs to a chassis where the preamplifier board is fixed.

After purging the inner chamber with Ar, it is valved off at atmospheric pressure and then closed in a vacuum shroud, which is then pumped down, prior to positioning the entire assembly on top of the dewar, see Fig.~\ref{fig:xx4}\,(middle), for about 2~h long cooldown to 88~K operating temperature.

%%%%%%%%%%%%%%%%%%%%%%%%%%%%%%%%%%%%%%%%%%%%%%%
%%%%%%%%%%%%%%%%%%%%%%%%%%%%%%%%%%%%%%%%%%%%%%%
%%%%%%%%%%%%%%%%%%%%%%%%%%%%%%%%%%%%%%%%%%%%%%%
%%%%%%%%%%%%%%%%%%%%%%%%%%%%%%%%%%%%%%%%%%%%%%%

\subsection{Relative wavelength-shifting efficiency measurement with ArGSet}\label{sec:xxx3}
\begin{figure}[h!]
        \centering\includegraphics[width=.85\linewidth]{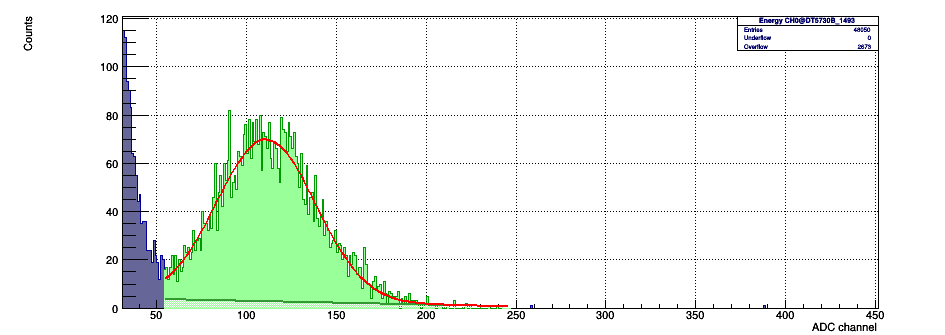}
        \includegraphics[width=.85\linewidth]{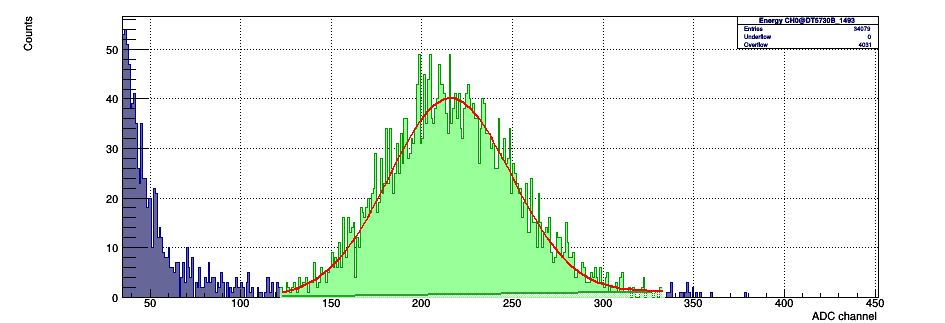}
    \caption{Energy spectra obtained with ArGSet for PEN (left) and TPB (right) materials using $^{241}$Am source.}
    \label{fig:main}
\end{figure}

During the commissioning two different substrates were tested. The wavelength-shifting performance was assessed for both PEN and TPB substrates so that their relative yield could be compared with the literature. In both runs the count rate was approx. 2 counts per second, consistent with the expected rate from the collimated $\alpha$ source. Fig.~\ref{fig:main} shows the spectra obtained for PEN (left) and TPB (right) materials using the previously described setup. No $\alpha$ peak in the spectrum and a negligible event rate was observed when taking data with vacuum (as opposed to gaseous argon) in the inner chamber, confirming that argon scintillation is the primary source of light.

The commissioning measurements revealed the wavelength shifting efficiency measurement, with 0.5 efficiency of PEN relative to TPB -- consistent with independent result for the same PEN batch~\cite{2pac}. Repeated cycle of measurements with the same sample, after warming up, purging and re-filling with argon (non-recirculated), gave a result consistent to 5\%. Analysis of the remaining systematic uncertainties is still ongoing.

\begin{figure}[h!] 
	\begin{center}  
		\includegraphics[height=2in,width=4in]{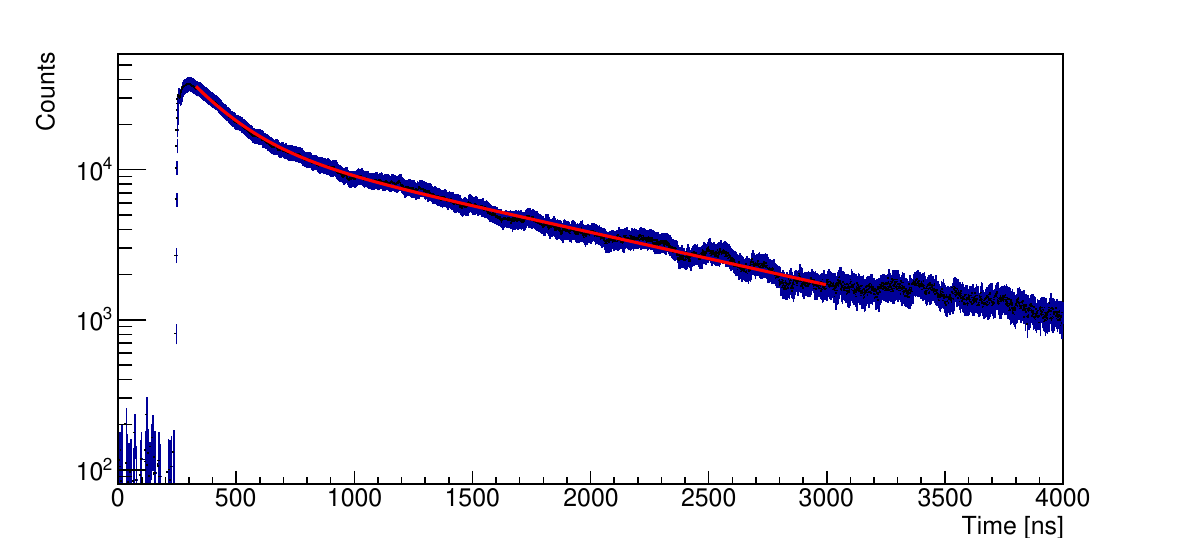}  
		\caption{Waveform used for extracting the triplet lifetime with a double exponential fit.}\label{fig:xx8}  
	\end{center}  
\end{figure}

Regarding the observed long lifetime the experimental result is shorter than the triplet lifetime for pure argon reported in literature. As the 6N Ar gas was delivered straight from the bottle and without purification in a hot getter, this can be explained by the presence of impurities known to be responsible for such behaviour. We plan to study such effects in more detail in the future with the help of a residual gas analyzer and, in particular, deliberately quench the triplet component with impurities in order to decouple WLS response from argon scintillation timing (based on the strategy from~\cite{Segreto}). Fig.~\ref{fig:xx8} shows a typical time spectrum used to extract the long lifetime constant (with the short time constant currently dominated by the SiPM response). 

Results obtained are summarized in Tab.~\ref{tab:xxx2}, with the literature results for the WLS efficiency and triplet lifetime constant taken from 2PAC~\cite{2pac} and MiniCLEAN~\cite{miniclean}, respectively. 
\begin{table}[h]
	\caption{Summary of results obtained in this work and data from previous works or literature.}
	\label{tab:xxx2}
	\smallskip
	\centering
	\begin{center}
		\begin{tabular}{  l  c  c }
			\hline\hline
			 & WLS efficiency (PEN/TPB)& Long lifetime $\mu$s \\[0.5ex] \hline
			Current work  & 0.5 & 1.24 \\ 
			Literature & 0.472$\pm$0.057 & 3.2  (pure gas)
			\\ \hline
		\end{tabular}
	\end{center}
	
\end{table}

\section{ArGSet upgrade for wavelength-shifting FAT-GEMs}
\begin{figure}[h!] 
\begin{center} 
\includegraphics[height=3in,width=6in]{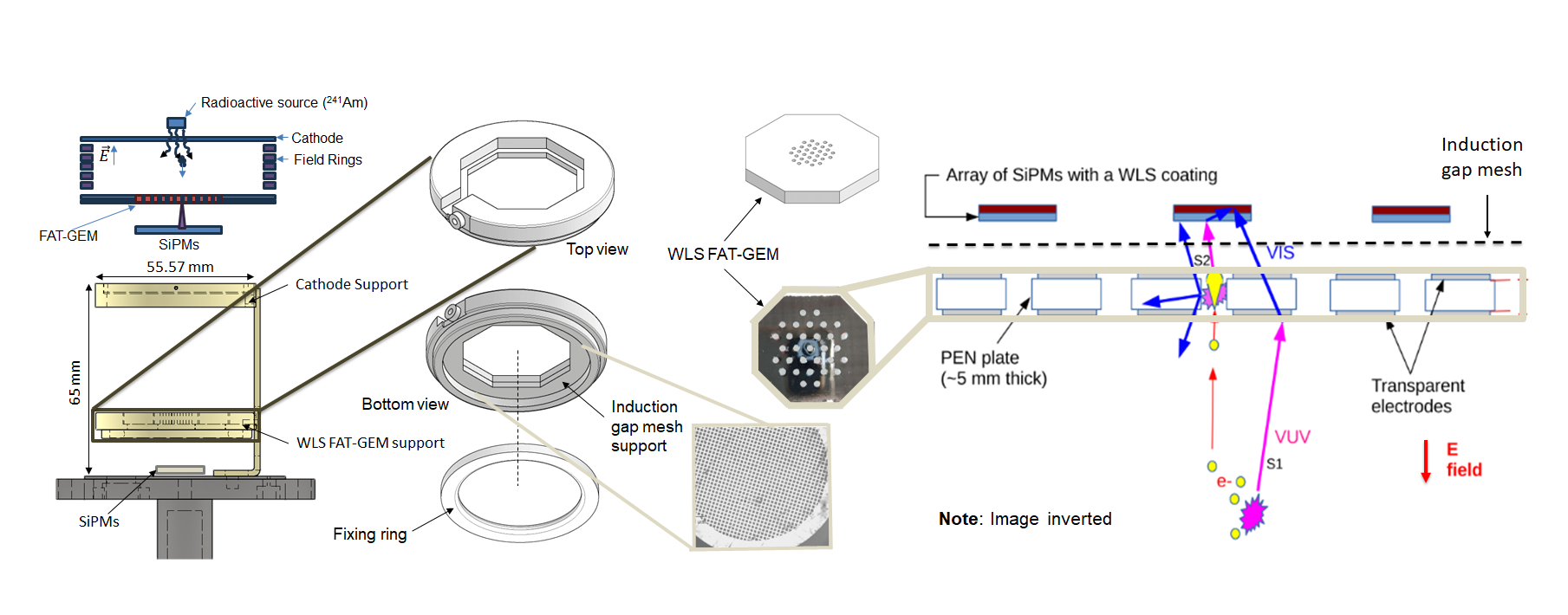}   
\caption{Schematic of the cryogenic setup adapted to test WLS FAT-GEM structures along with the working principle of the WLS FAT-GEMs. \label{fig:xx7}}  
\end{center}  
\end{figure}

The WLS FAT-GEM~\cite{wlsfatgem} presents several advantages when compared with state-of-the-art solutions. 
A schematic the polymeric (i.e. radiopure) FAT-GEM can be seen in Fig.~\ref{fig:xx7} (right) along with the working principle with a particular stress in the WLS and amplification mechanisms.

Confining S2 emission to the holes, makes dual-phase TPCs less sensitive to the liquid-gas interface uniformity, while suppressing electrostatic sagging and typical deformations on wire grids/meshes, making it versatile and easy to test in different configurations for improved light gain. 

Improving light collection is achieved by placing the amplification structure close to the photosensors, enabling wavelength-shifting (of S1 or S2 scintillation) in it and making it highly transparent to visible, already wavelength-shifted, light. Collection efficiency for S2 is increased due to the WLS effect of the holes’ walls.

Such structures further optimized for low optical losses can match or exceed the performance of state-of-the-art mesh solutions~\cite{wlsfatgem, leardini2022}.

%\subsection{ArGSet upgrade for WLS FAT-GEMs}
An extension to ArGSet’s current design is ongoing. The schematic of the planned extension to ArGSet's design can be seen in Fig.~\ref{fig:xx7} (left). This extension will allow testing different configurations and manufacturing processes used in the development of wavelength-shifting FAT-GEMs
~\cite{wlsfatgem, leardini2022}. Compared to other test stands~\cite{fatgem}, ArGSet will be unique in enabling such tests directly in cryogenic conditions.

The main features of the system are summarized below:
\begin{itemize}
	\item Modular structure made of cryogenic and radiopure materials (easy to change between WLS material studies and testing of FAT-GEMs);
	\item Independent biasing of internal structures (cathode, FAT-GEM and induction gap mesh) up to 10~kV;
	\item Charge and light readout;
	\item Allow for pressure (up to 5 bar) and temperature studies;
	\item Allow for varying the absorption/drift region from 5~mm up to 5~cm
\end{itemize}

Once ArGSet upgrade is finalized, data taking with different WLS FAT-GEM configurations will follow. The development of such system will allow the optimization of the WLS FAT-GEM performance with the objective of bridging the gap between these structures and state-of-the-art solutions with the ultimate goal of providing a scalable optical amplification structure for future dual-phase TPCs, contributing significantly to the development of the needed technology readiness for DM searches.
\section{Conclusion and Future work}\label{sec:xxx5}

We reported on a recently commissioned cryogenic setup that enables the study of the wavelength-shifting performance of candidate materials, and will also allow for the development of novel WLS FAT-GEMs. The cryogenic setup, ArGSet, was commissioned at NCBJ and Astrocent, and is now ready for extended measuring campaigns.

Future work planned includes:
\begin{itemize}
	\item Study the influence of impurities on the triplet lifetime constant;
	\item Analysis of systematic uncertainties;
	\item Quality control of $\sim$200~m$^2$ PEN batch for the DarkSide-20k veto;
	\item Survey of novel WLS materials;
	\item Measurement of Birks’ constant of chosen WLS;
	\item Operation with a nanosecond pulsed VUV light source;
    \item WLS FAT-GEM development and optimization.
\end{itemize}

\acknowledgments
This work was supported from the International Research Agenda Programme AstroCeNT (MAB/2018/7) funded by the Foundation for Polish Science from the European Regional Development Fund, and from the EU’s Horizon 2020 research and innovation programme under grant agreement No 952480 (DarkWave). Fruitful discussions with Diego González-Díaz (IGFAE/Univ. de Santiago de Compostela) on the system upgrade are gratefully acknowledged.

%% The Appendices part is started with the command \appendix;
%% appendix sections are then done as normal sections
%% \appendix

%% \section{}
%% \label{}

%% References
%%
%% Following citation commands can be used in the body text:
%% Usage of \cite is as follows:
%%   \cite{key}         ==>>  [#]
%%   \cite[chap. 2]{key} ==>> [#, chap. 2]
%%

%% References with bibTeX database:

% Bibliography

%% [A] Recommended: using JHEP.bst file
%% \bibliographystyle{JHEP}
%% \bibliography{biblio.bib}

%% or
%% [B] Manual formatting (see below)
%% (i) We suggest to always provide author, title and journal data or doi:
%% in short all the informations that clearly identify a document.
%% (ii) please avoid comments such as "For a review'', "For some examples",
%% "and references therein" or move them in the text. In general, please leave only references in the bibliography and move all
%% accessory text in footnotes.
%% (iii) Also, please have only one work for each \bibitem.

\end{document}